\documentclass[prl,twocolumn,showpacs,superscriptaddress,floatfix] {revtex4}
\usepackage{epsfig,color}
\usepackage{dcolumn}% Align table columns on decimal point
\newcommand{\p}{\mathbf{Q}}
\newcommand{\B}{\lambda}
\newcommand{\G}{\eta}

\newcommand{\beq}{\begin{equation}}
\newcommand{\eeq}{\end{equation}}
\newcommand{\bea}{\begin{eqnarray}}
\newcommand{\eea}{\end{eqnarray}}
\newcommand{\bwt}{\begin{widetext}}
\newcommand{\ewt}{\end{widetext}}

\begin{document}

\title{Non-analytic spin susceptibility of a nested Fermi liquid: the case of Fe-based pnictides}

\author{M.M.~Korshunov}
 \email{maxim@mpipks-dresden.mpg.de}
 \affiliation{Max-Planck-Institut f\"{u}r Physik komplexer Systeme, D-01187 Dresden, Germany}
 \affiliation{L. V. Kirensky Institute of Physics, Siberian Branch of Russian Academy of Sciences, 660036 Krasnoyarsk, Russia}
\author{I.~Eremin}
 \affiliation{Max-Planck-Institut f\"{u}r Physik komplexer Systeme, D-01187 Dresden, Germany}
 \affiliation{Institute f\"{u}r Mathematische und Theoretische Physik, TU Braunschweig, 38106 Braunschweig, Germany}
\author{D.V.~Efremov}
 \affiliation{Institut f\"ur Theoretische Physik, TU Dresden, 01062 Dresden, Germany}
\author{D.L.~Maslov}
 \affiliation{Department of Physics, University of Florida, Gainesville, Florida 32611, USA}
\author{A.V.~Chubukov}
 \affiliation{Department of Physics, University of Wisconsin-Madison, Madison,
Wisconsin 53706, USA}
\date{\today}

\begin{abstract}
We propose an explanation of the peculiar linear temperature dependence of the uniform spin susceptibility $\chi(T)$ in ferropnictides. We argue that the linear in $T$ term appears due to non-analytic temperature dependence of $\chi(T)$ in a two-dimensional Fermi liquid. We show that the prefactor of the $T$ term is expressed via the square of the spin-density-wave (SDW) amplitude connecting nested hole and electron pockets. Due to an incipient SDW instability, this amplitude is large, which, along with a small value of the Fermi energy, makes the $T$ dependence of $\chi(T)$ strong. We demonstrate that this mechanism is in quantitative agreement with the experiment.
\end{abstract}


\maketitle

\textit{Introduction.} The key hypothesis of the Fermi liquid (FL)
theory is that a system of strongly interacting fermions can be
considered effectively as a gas of weakly interacting quasiparticles. In the absence of residual interaction between quasiparticles, the static uniform spin susceptibility, $\chi(T)$, and the specific heat coefficient, $\gamma(T)$, are finite at $T=0$ and obey quadratic dependencies on $T$ at low temperatures. The effect of residual interactions on $\chi(T)$ and $\gamma(T)$ has been studied intensively in recent years \cite{BelitzWoelfle}, with the key result that in two dimensions (2D) both $\gamma(T)$ and $\chi(T)$ are linear rather than quadratic in $T$ \cite{MaslovChubukov03,Betouras2005,MaslovChubukov2008}.

Theory predicts that the behavior of $\gamma(T)$ is universal in a sense that the (negative) slope is given by the square of the backscattering amplitude \cite{cmgg}. This linear decrease has  been observed in monolayers of $^3$He \cite{saunders}. On the contrary, a linear in $T$ term in the spin susceptibility is not universal and can be of either sign \cite{finn,MaslovChubukov2008}, causing uncertainty in the interpretation of the experiments on semiconductor heterostructures \cite{reznikov}.

Recently, a pronounced linear temperature dependence of the uniform susceptibility has been observed in high-$T_c$ superconductors with
iron-based layered structure \cite{Wu2008,Yan2008,Klingeler2008,Wang2008}. It extends from temperatures above either SDW or superconducting transitions up to 500-700K with almost doping-independent slope. The $T$ dependence is quite strong: $\chi(T)$ increases roughly by a factor of 2 between 200K and 700K. An explanation of this behavior based on the $J_1-J_2$ model of localized spins has been proposed in Ref.~\onlinecite{d_lee}; however, given that the linear $T$ dependence persists up to large dopings, where local probes, such as nuclear magnetic resonance (NMR) and $\mu$SR, do not see localized moments \cite{Nakai_NMR,Klauss2008}, this explanation is questionable.

The itinerant character of Fe-pnictides is suggested by the agreement between the band structure obtained in \textit{ab initio} calculations \cite{LDA} and observed in de Haas-van Alphen and angle-resolved photoemission spectroscopy (ARPES) experiments \cite{Coldea2008,Lu2008,Liu2008,Liu2008_2,ding_latest}. It is firmly established by now that the Fermi surface (FS) of Fe-pnictides consists of two small hole pockets near $(0,0)$ and two electron pockets near $(\pi,\pi)$ points of the folded Brillouin zone. In such a system, an obvious origin of the SDW order is a logarithmic divergence of a particle-hole vertex involving states on nested parts of the FS \cite{LDA,Dong,Tesanovic,Chubukov2008,d_h_lee,Korshunov2008}.

In this Letter, we propose an explanation of the experimental $T$-dependence of spin susceptibility based on the itinerant picture. We argue that the origin of the linear increase of $\chi(T)$ with temperature in ferropnictides is the same as in a 2D FL. Furthermore, we show that this behavior is universal for FLs with strong $(\pi,\pi)$ SDW fluctuations, namely, the slope of the linear in $T$-dependence is determined by the square of the SDW amplitude with nesting momentum $\p=(\pi,\pi)$. This amplitude is large which, along with a small value of the Fermi energy $\varepsilon_F$, amplifies the $T$ dependence of $\chi (T)$. Choosing the SDW coupling to reproduce the observed SDW transition temperature $T_N$ at zero doping, we find a good agreement between calculated and measured slopes of $\chi (T)$. The doping dependence of the slope also agrees with the data. We view this agreement as a strong indication that the linear temperature dependence of $\chi(T)$ in pnictides \cite{Wu2008,Yan2008,Klingeler2008,Wang2008} is in fact the first unambiguous observation of a non-analytic behavior of the 2D spin susceptibility. Besides being fundamentally important on its own right, this observation also strengthens the case for the itinerant scenario for Fe-pnictides.

\textit{Theory.} We consider a two-band model of interacting fermions occupying the electron and hole FSs:
\begin{eqnarray}
\label{eqH}
H&=& \sum_{\mathbf{k}, \sigma} \left[ \varepsilon^c_\mathbf{k} c_{\mathbf{k} \sigma}^\dag c_{\mathbf{k} \sigma}
 + \varepsilon^f_\mathbf{k} f_{\mathbf{k} \sigma}^\dag f_{\mathbf{k} \sigma} \right]
 + \sum_{\mathbf{p}_i, \sigma, \sigma'} H_{\mathrm{int}},
 \\
H_{\mathrm{int}}&=&u_1 c^\dag_{\mathbf{p}_3 \sigma} f^\dag_{\mathbf{p}_4 \sigma'} f_{\mathbf{p}_2 \sigma'} c_{\mathbf{p}_1 \sigma}
 + u_2 f^\dag_{\mathbf{p}_3 \sigma} c^\dag_{\mathbf{p}_4 \sigma'} f_{\mathbf{p}_2 \sigma'} c_{\mathbf{p}_1 \sigma} \nonumber \\
&+& \frac{u_3}{2} \left[ f^\dag_{\mathbf{p}_3 \sigma} f^\dag_{\mathbf{p}_4 \sigma'}
 c_{\mathbf{p}_2 \sigma'} c_{\mathbf{p}_1 \sigma} + H.c \right] \nonumber \\
&+& \frac{u_4}{2} f^\dag_{\mathbf{p}_3 \sigma} f^\dag_{\mathbf{p}_4 \sigma'} f_{\mathbf{p}_2 \sigma'} f_{\mathbf{p}_1 \sigma}
 + \frac{u_5}{2} c^\dag_{\mathbf{p}_3 \sigma} c^\dag_{\mathbf{p}_4 \sigma'} c_{\mathbf{p}_2 \sigma'} c_{\mathbf{p}_1 \sigma}. \nonumber
\end{eqnarray}
Here, $c_{\mathbf{k} \sigma}$ ($f_{\mathbf{k} \sigma}$) is the annihilation operator for a hole (electron) with momentum $\mathbf{k}$ and spin $\sigma$ (for an electron, $\mathbf{k}$ is measured from the $(\pi,\pi)$-point), $\varepsilon^c_\mathbf{k} $ and $\varepsilon^f_\mathbf{k}=-\varepsilon^c_\mathbf{k} + 2\mu$ represent
single-particle dispersions, and $\mu$ measures a deviation from perfect nesting. Models of this type were considered in the past in the context of an ``excitonic insulator'' \cite{ED}.

We assume that each of the electron and hole FSs is doubly degenerate. The terms with $u_4$ and $u_5$ are intra-band interactions, the terms with $u_1$ and $u_2$ are inter-band interactions with momentum transfer $0$ and $\p$, respectively, and the term with $u_3$ is the inter-band pair hopping. All couplings flow from their initial values at energies of order of the bandwidth to renormalized values at $\varepsilon_F$ \cite{Chubukov2008,d_h_lee}. We assume that this renormalization is already included into Eq.~(\ref{eqH}) and analyze the behavior of the system at energies below ${\varepsilon}_F$, where the actual values of momenta become relevant.

\begin{figure}
\includegraphics[angle=0,width=0.99\columnwidth]{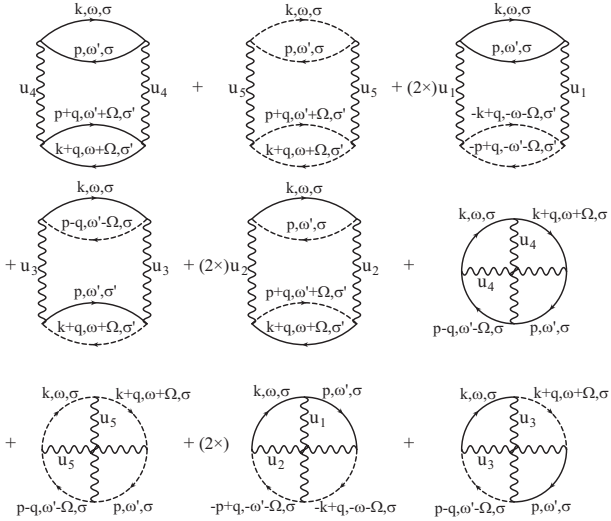}
\caption{Second-order diagrams for the thermodynamic potential. Solid and dashed lines correspond to $f$-fermions (electrons) and to $c$-fermions (holes), respectively.}
\label{fig1}
\end{figure}
We first obtain the linear-in-$T$ contribution to the spin susceptibility, $\delta\chi(T)$, to second order in the interaction and then show that the prefactor of the $T$ term in $\delta\chi(T)$ is expressed via the SDW vertex to all orders in the interaction.

Fig.~\ref{fig1} depicts all topologically inequivalent second-order diagrams for the thermodynamic potential, $\Phi(T,H)$, describing both intra- and inter-band processes. Each of the diagrams contains the object
\begin{eqnarray}
\varphi(T,H) &=&  T^3 \sum_{\Omega} \int \frac{d^2\mathbf{q}}{(2\pi)^2} \sum_{\sigma,\sigma'} \sum_{\omega,\omega'} \int \frac{d^2\mathbf{k}
d^2\mathbf{p}}{(2\pi)^4} \nonumber \\
 &\times& G^{\sigma}_{\mathbf{k},\omega} G^{\sigma}_{\mathbf{p},\omega'}
G^{\sigma'}_{\mathbf{p+q},\omega'+\Omega} G^{\sigma'}_{\mathbf{k+q},\omega+\Omega},
\label{eqPhi}
\end{eqnarray}
where $G^{\sigma}_{\mathbf{k},\omega}$ is a Matsubara Green's function of either $c$ or $f$ fermion with momentum $\mathbf{k}$ and spin-$\sigma$ in a magnetic field $H$. As in an ordinary FL, the non-analytic $H^2 T$ term in $\varphi(T,H)$ comes from a dynamic Kohn anomaly, i.e. from diagrams with momentum $2k_F$ carried by the interaction lines \cite{MaslovChubukov2008,Betouras2005}. It is, however, more convenient to re-express the result via the polarization bubbles with small rather than $2k_F$ momenta. The Green functions in Eq.~(\ref{eqPhi}) can be combined in two different ways: either as
$\Pi^{\sigma\sigma}_{\mathbf{q},\Omega} \Pi^{\sigma' \sigma'}_{\mathbf{q},\Omega}$, or as $\Pi^{\sigma \sigma'}_{\mathbf{q},\Omega} \Pi^{\sigma \sigma'}_{\mathbf{q},\Omega}$, where $\Pi^{\sigma \sigma'}_{\mathbf{q},\Omega} =T \sum_{\omega} \int \frac{d^2\mathbf{k}}{(2\pi)^2} G^{\sigma}_{\mathbf{k},\omega} G^{\sigma'}_{\mathbf{k+q},\omega+\Omega}$ is a polarization
bubble of the $cc$, $ff$, or $cf$ type with small momenta $q \ll k_F$. A bubble depends on the magnetic field if the Zeeman energies of two fermions add up. One can readily show that the $cc$ and $ff$ bubbles depend on the field via ${\Pi^{\uparrow\downarrow}_{\mathbf{q},\Omega}}$, while the field enters the $cf$ bubble through ${\Pi^{\uparrow\uparrow}_{\mathbf{q},\Omega}}$ and ${\Pi^{\downarrow\downarrow}_{\mathbf{q},\Omega}}$ terms. Evaluating individual diagrams, we find that each of them can be expressed
as a product of two dynamic spin up/down bubbles:
\begin{equation}
\Pi_{\mathbf{q},\Omega} = \frac{m}{2\pi} \frac{|\Omega|}{\sqrt{(\Omega-2
\mathrm{i} \mu_B H - 2 \mathrm{i} \mu^*)^2 + (v_{F} q)^2}},
\label{eqPi}
\end{equation}
where $\mu^*=0$ for intra-band scattering and $\mu^*=\mu$ for inter-band scattering. The rest of the calculations proceeds in the same way as for an ordinary FL \cite{Betouras2005,MaslovChubukov2008}. In short, one integrates $\Pi^2_{\mathbf{q},\Omega}$ over $\mathbf{q}$ first, replaces the Matsubara sum by a contour integral, differentiates the result twice with respect to $H$, sets $H=0$, and obtains the $O(T)$ terms in $\chi$. The first two diagrams in Fig.~\ref{fig1} describe intra-band processes and give the same results for $\chi(T)$ as in Refs.~\cite{MaslovChubukov2008,Betouras2005}:
$\delta\chi_{1,2}(T)=\B u_{4,5}^2 T$, where $\B = 4 \chi_0 N_F^2/(2\varepsilon_F)$, $\varepsilon_F=v_F k_F/2$, $\chi_0=2\mu_B^2 N_F$ is the Pauli susceptibility per one sheet of the FS, and $N_F=m/2\pi$ is the density of states. A factor of $4$ in $\B$ results from the double degeneracy of electron and hole bands. The third through fifth diagrams involve inter-band scattering and yield $\delta\chi_3(T)=2 \B u_1^2 T \G(\mu/T)$, $\delta\chi_4(T)= \B u_3^2 T \G(\mu/T)$, and $\delta\chi_5(T)=2 \B u_2^2 T \G(\mu/T)$, respectively. Here, $\G(x)=2x \coth{x} - x^2/\sinh^2{x} - 2x$. The third and forth diagrams give finite contributions if $\sigma=\sigma'$, while the fifth diagram contributes if $\sigma=-\sigma'$. The sixth and seventh diagrams do not contribute to
$\delta\chi(T)$, and the remaining two give $\delta\chi_8(T)=-2 \B u_1 u_2 T \G(\mu/T)$ and $\delta\chi_9(T)=- \B u_3^2 T \G(\mu/T)$.

Combining all diagrams, we obtain
\begin{equation}
\delta\chi(T)= \B T \left[u_4^2+u_5^2+2\left(u_1^2+u_2^2-u_1 u_2\right) \G(\mu/T)\right].
\label{eqChi}
\end{equation}
The first two terms are the contributions from intra-band processes --
they are the same as in an ordinary FL. The rest of the terms correspond to inter-band processes, specific to the electronic structure of nested  ferropnictides. For $T \gg \mu$ (perfect nesting), $\G(\mu/T) \approx 1$ and both intra- and inter-band processes contribute to the linear term in the spin susceptibility. In the opposite limit of $T \ll \mu$ (poor nesting), $\G(\mu/T) \sim \exp(-2\mu/T)$ and the inter-band contribution to $\chi(T)$ is suppressed exponentially. In the intermediate regime $T \sim \mu$, $\G(\mu/T)$ is non-monotonic. Note that the pair-hopping term $u_3$, which gives rise to an attraction in the extended $s$-wave ($s_\pm$) pairing channel, does not contribute to the $T$ term in $\chi(T)$.

The coupling constants in Eq.~(\ref{eqChi}) are interactions at the scale
of the Fermi energy. These couplings are already renormalized from their bare values at the scale of the bandwidth \cite{Chubukov2008} by parquet
renormalization group (RG), in such a way that $u_4$, $u_5$, and $u_1$ flow to the same value, while $u_2$ flows to a smaller value. The renormalized coupling at the scale of $\varepsilon_F$ depend only weakly on the incoming and transferred momenta. Further renormalization \textit{below} $\varepsilon_F$ differentiate between $u_i$ with different
transformed and incoming momentum. Such renormalization is particularly
relevant for our system as the coupling $u_1$, which corresponds to the scattering process, $u_1\left(\mathbf{k},\mathbf{p+q};\mathbf{p},\mathbf{k+q}\right)
c^\dagger_\mathbf{p}f^\dagger_\mathbf{k+q} f_\mathbf{p+q}c_\mathbf{k}$,
diverges at the onset of the SDW instability for $\mathbf{q}=0$
(we recall that momenta for $f$-fermions are measured from $\p$, so that the momentum transfer between $c$ and $f$ fermions is actually $\mathbf{q}=\p$). The singular vertex is $u_1(\mathbf{k},\mathbf{p};\mathbf{p},\mathbf{k})$, which means that electrons and holes swap their respective momenta. Note that the divergence occurs for any angle between $\mathbf{k}$ and $\mathbf{p}$. At weak coupling, the enhancement of $u_1$ is confined to $q \ll k_F$, while for a generic $q \sim k_F$ the coupling $u_1$ retains its bare value.

We now show that the $u_1^2$ term in $\chi(T) \propto T$ is the same
coupling that diverges at the SDW instability. To see this, we first note that the fermionic momenta in diagrams for $\Phi(T,H)$ are constrained by two requirements: i) the momentum transfers are near $2k_F$ and ii) all four momenta are near the FS. For the third diagram in Fig.~\ref{fig1},
this implies that $\mathbf{p} \approx -\mathbf{k}$, $|\mathbf{k}| \approx k_F$, while $q$ is small ($\sim T/v_F$). Therefore, the vertex in this diagram is $u_1(\mathbf{k},-\mathbf{k}; \mathbf{-k},\mathbf{k})$. This is an analog of the backscattering amplitude in a 2D FL. The vertex $u_1(\mathbf{k},-\mathbf{k};\mathbf{-k},\mathbf{k})$ is a special
case of the SDW vertex $u_1(\mathbf{k},\mathbf{p};\mathbf{p},\mathbf{k})$ for $\mathbf{p}=-\mathbf{k}$.

\begin{figure}
\includegraphics[angle=0,width=1\columnwidth]{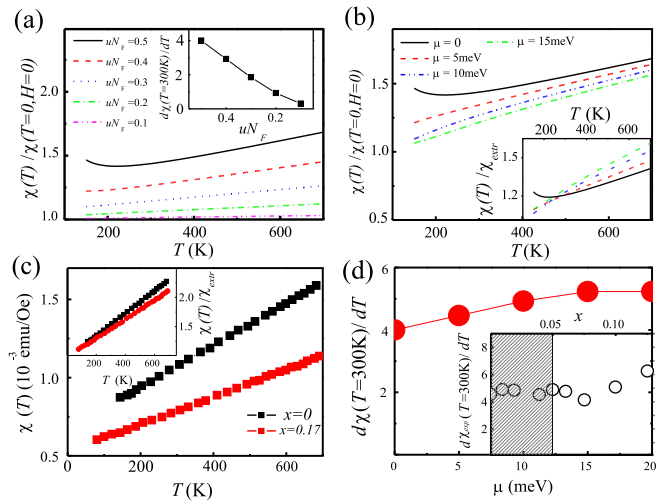}
\caption{(a) $T$ dependence of the spin susceptibility $\chi(T)$, as given by Eq.~(\ref{eqChi_new_last}), for perfect nesting ($\mu=0$) and for a range of couplings $uN_F$, as indicated in the plot. In a calculation, $\mathrm{max}(T,\mu)$ in Eq.~(\ref{eqChi_new_last}) is approximated by $\sqrt{T^2+\mu^2}$. Inset: Calculated slope $d\chi(T)/dT$ as a function of $uN_F$ at $T=300$K (chosen to match Ref.~\onlinecite{Klingeler2008}). (b) Calculated $\chi(T)$ for $uN_F=0.5$ for a range of $\mu$, as shown in the plot. Inset: Same curves as in (a) normalized by $\chi(0)_{extr}$, obtained by extrapolating $\chi(T)$ down to $T=0$. (c) $\chi(T)$ in BaFe$_{2-x}$Co$_x$As$_2$ from Ref.~\onlinecite{Wang2008}. Inset: same data as in the main panel normalized by $\chi(0)$. (d) Calculated slope $d\chi(T=300\mathrm{K})/dT$ as function of $\mu$. Inset: measured slope as a function of doping (from Ref.~\onlinecite{Klingeler2008}).}
\label{fig2}
\end{figure}
Next, we consider higher-order diagrams. They can be separated into two classes. In diagrams of the first class, one obtains the non-analyticity by keeping only two dynamic bubbles $\Pi_{\mathbf{q},\Omega}$ and lumping the rest of the diagram into renormalization of the static scattering amplitude. In particular, the backscattering amplitude $u_1(\mathbf{k},\mathbf{-k};-\mathbf{k},\mathbf{k})$ is renormalized into
an effective coupling $u^{\mathrm{eff}}_1$, which diverges at the SDW instability. Singular renormalizations of $u_1$ form ladder series which is summed into
\begin{equation}
u_1^{\mathrm{eff}} = \frac{u_1}{1-u_1 N_F \ln \frac{\varepsilon_F}{\mathrm{max}(T,\mu)}}.
\label{ueff}
\end{equation}
The second class of higher-order diagrams contain three and more dynamic
bubbles. Terms of order of $u_1^3$ etc. are \textit{not} expressed in terms of backscattering amplitude but rather contain $u_1$ with typical $q$ of order $k_F$. In this range of momenta, $u_1$ is not enhanced by SDW fluctuations and remains small at weak coupling.

Neglecting $u_2$ and setting $u_1=u_4=u_5 \equiv u>0$, we finally obtain
\begin{equation}
\delta\chi(T) \approx \frac{8 \left(u N_{F}\right)^2 \chi_0}{v_F k_F} T \left[1+ \frac{\G(\mu/T)}{\left(1-uN_F\ln{\frac{\varepsilon_F}{\mathrm{max}(T,\mu)}}\right)^2} \right].
\label{eqChi_new_last}
\end{equation}
We see that the full result for $\delta\chi(T)$ is obtained from the second-order expression by replacing $u_1=u$ by the exact SDW amplitude $u_1^{\mathrm{eff}}$, given by Eq.~(\ref{ueff}). This is similar to the result for the specific heat \cite{cmgg}, but differs from the result for $\delta\chi(T)$ in an ordinary 2D FL, where the full $\delta\chi(T)$ is not expressed via the backscattering amplitude \cite{MaslovChubukov2008,finn}. This difference can be traced down to the symmetry between the particle-particle (Cooper) channel in an ordinary FL and the particle-hole channel in a nested FL. Indeed, the backscattering contribution to both $\gamma(T)$ and $\delta\chi(T)$ for the ordinary case undergoes logarithmic renormalization in the Cooper channel. However, the Cooper ladder for the ordinary case is identical to the particle-hole ladder for the nested case, except for that the sign of the interaction is reversed, i.e. the SDW instability for $u>0$ for the nested case is related to the Cooper instability for $u<0$ for the ordinary case. In both cases, there are also non-backscattering contributions $\delta\chi(T)$. For the ordinary case, the backscattering term is reduced by Cooper renormalization, and non-backscattering contributions play the dominant role. For the nested case, SDW renormalization enhances the backscattering term, and other contributions can be ignored. On the other hand, the Cooper channel composed of electrons and holes for the nested case is equivalent to the particle-hole channel for an ordinary case and, therefore, is not logarithmically divergent.

\textit{Comparison with experiments.} We now apply Eq.~(\ref{eqChi_new_last}) to ferropnictides. The experimental results for $\chi(T)$ in BaFe$_{2-x}$Co$_x$As$_2$ \cite{Wang2008} are shown in panel (c) of Fig.~\ref{fig2}. From the data, we estimate the slope of the $T$ dependence as $\left(\chi(700\mathrm{K})/\chi(0)\right)_{\mathrm{exp}} \approx 2$, where $\chi(0)$ is obtained by extrapolating $\chi(T)$ to $T=0$ (theoretically, $\chi(0) \approx 4\chi_0$). Taking $v_F=0.45$eV$\cdot\AA$ and $k_F \approx 0.16\AA^{-1}$ from
the ARPES data \cite{ding_latest}, we obtain $\varepsilon_F \sim
0.04$ eV \cite{comment_EF}. The only unknown parameter of the theory -- the dimensionless coupling constant $uN_F$ -- is fixed by requiring that the SDW vertex $u_1^{\mathrm{eff}}$ increases upon approaching
$T_N$ ($T_N=140$K at zero doping). As Eq.~(\ref{ueff}) is an approximate one-loop formula, we set a criterium that $u_{eff}$ increases by a factor of $2$ at $T_N$. This yields $uN_F \approx 0.5$. We then find $\left(\chi(T=700\mathrm{K})/\chi(0)\right)_{\mathrm{theor}} \approx 1.7$, which is quite close to the experimental $\left(\chi(700\mathrm{K})/\chi(0)\right)_{\mathrm{exp}} \approx 2$.
A more detailed comparison between the experiments and Eq.~(\ref{eqChi_new_last}) is presented in Fig.~\ref{fig2}, where we also
show the dependencies of the slope on $uN_F$ and $\mu$. We find quite a good agreement with the experimental data.

\textit{Conclusion.} To summarize, we analyzed a non-analytic, linear in $T$ term in the spin susceptibility of a 2D Fermi liquid with nested electron and hole pockets of the Fermi surface. We found that the prefactor of the $T$-term contains the same inter-band coupling $u_1^{\mathrm{eff}}$, which is enhanced by SDW fluctuations. These results describe quantitatively a strong temperature dependence of the spin susceptibility in ferropnictides, observed in a number of recent experiments. An immediate consequence of the proposed mechanism is that $\chi$ should exhibit equally strong linear dependencies on the magnetic field and on the wave number \cite{MaslovChubukov2008}. We suggest to perform these measurements as a crucial test for the origin of the observed effect.

We thank R. Klingeler for useful discussions. M.M.K.
acknowledges support from RFBR 07-02-00226, OFN RAS program on ``Strong electronic correlations'', and RAS program on ``Low temperature quantum phenomena''. I.E. acknowledges support from Asian-Pacific Center for Theoretical Physics. D.L.M. acknowledges support from Laboratoire de Physique des Solides, Universit{\'e} Paris-Sud (France) and RTRA Triangle de la Physique. A.V.C. acknowledges support from NSF-DMR 0604406.


\begin{thebibliography}{99}
\bibitem{comment_EF} For a generic dispersion, the Fermi energy, $\varepsilon_F$, defined in this way does not correspond to the energy scale separating the regimes of Fermi and Boltzmann statistics. The latter occurs at $T \sim T^* \equiv \sqrt{N_F/N_F''}$. It can be shown that higher than linear terms in $\delta\chi(T)$ become important also at $T \sim T^*$. We assume that $T^*$ in pnictides is high enough so that temperatures of interest for this study correspond to Fermi statistics and linear scaling of $\delta\chi(T)$.
\end{thebibliography}
\end{document}